\documentstyle[12pt]{article}
\makeatletter
\@addtoreset{equation}{section}
\makeatother
\renewcommand{\theequation}{\thesection.\arabic{equation}}
\newcommand {\equ}[1] {(\ref{#1})}
\advance\hoffset by -7mm
\renewcommand{\title}[1]{\null\vspace{25mm}

\noindent{\Large{\bf #1}}\vspace{10mm}

\noindent {\large }}
\newcommand{\authors}[1]{\noindent{\large #1}\vspace{5mm}

}
\newcommand{\address}[1]{\noindent #1\vspace{10mm}

}
\renewcommand{\abstract}[1]{\vspace{17mm}

\noindent{\small{\em Abstract.} #1}\vspace{2mm}

}
\newcommand{\journal}[4]{{\em #1~}#2\,(19#3)\,#4;}
\newcommand{\aihp}{\journal {Ann. Inst. Henri Poincar\'e}}

\newcommand{\jmp}{\journal {J. Math. Phys.}}

\newcommand{\cmp}{\journal {Comm. Math. Phys.}}

\newcommand{\np}{\journal {Nucl. Phys.}}
\newcommand{\pl}{\journal {Phys. Lett.}}

\newcommand{\annp}{\journal {Ann. Phys. (N.Y.)}}

\def\f{\frac}
\def\be{\begin{equation}}
\def\ee{\end{equation}}
\def\bea{\begin{eqnarray}}
\def\eea{\end{eqnarray}}
\def\bean{\begin{eqnarray*}}
\def\eean{\end{eqnarray*}}
\def\ba{\begin{array}} \def\ea{\end{array}}
\newcommand{\eq}{\begin{equation}}
\newcommand{\eqn}[1]{\label{#1}\end{equation}}
\newcommand{\lp}{\left(}\newcommand{\rp}{\right)}

\newcommand{\lac}{\left\{}\newcommand{\rac}{\right\}}
\newcommand{\eqa}{\begin{eqnarray}}
\newcommand{\eqan}[1]{\label{#1}\end{eqnarray}}

\def\6{\partial}  
 \def\d{\delta}  \def\e{\epsilon}
 \def\h{\eta} 
  \def\l{\lambda}
\def\m{\mu} \def\n{\nu} \def\x{\xi}

  \def\O{\Omega}

\def\ti{\tilde} \def\wti{\widetilde}
\def\non{\nonumber\\}
\def\={\!\!\!&=&\!\!\!}
\def\n={\!\!\!&\not=&\!\!\!}
\def\+{\!\!\!&&\!\!\!+~}
\def\-{\!\!\!&&\!\!\!-~}

\renewcommand{\AA}{{\cal A}}

\newcommand{\FF}{{\cal F}}

\newcommand{\LL}{{\cal L}}

\newcommand{\NN}{{\cal N}}

\begin{document}

\setcounter{page}{0}
\thispagestyle{empty}\hspace*{\fill} REF. TUW 94-17

\title{BRST cohomology of Yang-Mills gauge fields in the presence
       of gravity in Ashtekar variables}

\authors{O. Moritsch\footnote{Work supported in part by the
         ``Fonds zur F\"orderung der Wissenschaftlichen Forschung''
         under Contract Grant Number P9116-PHY.},
         M. Schweda,
         T. Sommer\footnote{Work supported in part by the
         ``Fonds zur F\"orderung der Wissenschaftlichen Forschung''
         under Contract Grant Number P10268-PHY.},
         L. T\u{a}taru\footnote{Supported by the
         ``\"{O}sterreichisches Bundesministerium f\"{u}r
         Wissenschaft und  Forschung''.}
         \footnote{Permanent address Dept. Theor. Phys.,
         University of Cluj, Romania.}
         and H. Zerrouki\footnote{Work supported in part by the
         ``Fonds zur F\"orderung der Wissenschaftlichen Forschung''
         under Contract Grant Number P10268-PHY.}}
\address{Institut f\"ur Theoretische Physik,
         Technische Universit\"at Wien\\
         Wiedner Hauptstra\ss e 8-10, A-1040 Wien (Austria)}
\leftline{September 1994}
\abstract{
The BRST transformations for the Yang-Mills gauge fields in the
presence of gravity described by Ashtekar variables are obtained by
using the so-called Maurer-Cartan horizontality conditions.
The BRST cohomology group expressed by the Wess-Zumino consistency
condition is solved with the help of an operator $\delta$ introduced
by S.P. Sorella which in our case has a very simple form and
generates, together with the differential $d$ and the BRST operator
$s$, a simpler algebra than in the pure Yang-Mills theory.
In this way we shall find the Yang-Mills Lagrangians,
the Chern-Simons terms and the gauge anomalies.
}


\newpage

\section{Introduction}

\setcounter{equation}{0}

Gauge fields play a very important role in all theories which
describe fundamental interactions \cite{ym}.
Electroweak theory and quantum chromodynamics (QCD) are
examples of Yang-Mills theories.
The most efficient way to study the quantization and the
renormalization of such theories is given by
the introduction of BRST tranformations and by calculation of
the so-called BRST cohomology. In despite of the fact that
gravity could be introduced as a gauge theory associated
with local Lorentz invariance \cite{u}, its action
has a different structure and it is difficult to connect it
to a special form of Yang-Mills theory, known as the topological
quantum field theory (TQFT) \cite{w} (see also \cite{bbrt}).
It is necessary to desribe the propagating degree of freedom using
some variables that are related naturally to those employed in TQFT.
In particular, we should require that these variables are suitable
for implementing both diffeomorphism and gauge invariance.
In this respect, the variables introduced by Ashtekar \cite{ash}
satisfy these requirements.
In place of the metric of general relativity, the classical Ashtekar
variables \cite{ash,sp,rov,cs,cs1}, corresponding to specific
Einstein manifolds, consist of simple $SO(3)$-gauge fields
satisfying self-dual conditions. We shall analyze the BRST
symmetry \cite{brs} of 4-dimensional gravity,
described by the Ashtekar variables, coupled to the Yang-Mills
fields, i.e. the case of Yang-Mills gauge fields in a curved
spacetime, and we shall determine the solution of the descent
equations identifying the BRST invariants constructed out of them.
These quantities can be used to describe the diffeomorphism and
gauge invariance of the observables of the theory, and to deliver
characterizations of different structures of the manifold or of
the space of Yang-Mills fields.

The BRST invariant Lagrangians, possible anomalies and Schwinger
terms are nontrivial solutions of the Wess-Zumino consistency
condition \cite{wz}
\be
\label{CE}
s\Delta=0~~~,~~~\Delta\neq s\widehat{\Delta} \ ,
\ee
where  $\Delta$ and
$\widehat{\Delta}$ are integrated local polynomials in the fields
and their derivatives and $s$ is the nilpotent BRST operator.
In particular, the BRST formalism allows the characterization of
classical actions and anomalies as BRST invariant functionals.
Especially, an action is a BRST invariant functional with ghost
number zero, an anomaly corresponds to a BRST invariant functional
with ghost number one and a Schwinger term to such one with
ghost number two.

Setting $\Delta=\int{\cal B}$, condition (\ref{CE}) translates into
the local equation
\be
s{\cal B}+d{\cal Q}=0 \ ,
\label{des}
\ee
where ${\cal Q}$ is some local polynomial
in the fields and their derivatives and $d=dx^\mu\partial_\mu$
represents the exterior spacetime differential which, together
with the BRST operator $s$, obeys the BRST algebra
\be
s^2=d^2=sd+ds=0 \ .
\label{nilpot}
\ee
The local term ${\cal B}$ is called nontrivial if
\be
{\cal B}\neq s\widehat{\cal B}+d\widehat{\cal Q} \ ,
\label{nontr}
\ee
with $\widehat{\cal B}$ and $\widehat{\cal Q}$ local polynomials.
In this case the integral of ${\cal B}$ on spacetime,
$\int{\cal B}$, is a representative of a cohomology class
of the BRST operator $s$. The integrated cohomology problem
(\ref{CE}) is equivalent to the local one
\be
sa=0~~~,~~~a\neq s\hat{a} \ ,
\label{CEL}
\ee
where $a$ is a local polynomial in the fields and their derivatives.
The local equation (\ref{des}) could be solved
in a simple way by using the operator $\delta$ introduced
by S.P. Sorella \cite{sor} which obeys the decomposition of the
exterior spacetime derivative as a BRST commutator
\be
d=-[s,\delta] \ .
\label{sor}
\ee
Actually, the decomposition (\ref{sor}) represents
one of the crucial features of the topological field theories
\cite{gms,br} and the bosonic string and superstring in the
Beltrami and super-Beltrami parametrization \cite{wss,boss}.
As shown in \cite{mss,diss}, the eq.(\ref{sor}) allows a
cohomological interpretation of the cosmological constant, of
Lagrangians for pure Einstein gravity and generalizations including
also torsion, as well as gravitational Chern-Simons terms and
anomalies.
Due to the existence of the decomposition the study
of the cohomology of $s$ modulo $d$ (\ref{des}) is essentially
reduced to the study of the local cohomology of
$s$ (\ref{CEL}) which in turn can be systematically analyzed by
using the powerful techniques of spectral sequences \cite{dix,ban}.
In fact, as proven in \cite{st}, the solution obtained by utilizing
the decomposition (\ref{sor}) is completely equivalent to that
based on the {\em Russian formula}
\cite{tm1,sto,witten,ba1,dvio,gins,ton1,stora,brandt},
i.e. they differ only by trivial cocycles.

In a first step we will demonstrate
that the BRST transformations for the Yang-Mills fields as well as
the Ashtekar variables describing gravity, can be derived from
Maurer-Cartan horizontality conditions \cite{tm1,ba1,tm2,bb}.
After that we will prove that the decomposition (\ref{sor})
can be succesfully extended to Yang-Mills gauge theory coupled
to the Ashtekar variables, i.e. to the gravitational fields.
Finally, we will see that the operator
$\delta$ gives an elegant and straightforward way to classify the
cohomology classes of the full BRST operator.
As shown in \cite{bmsstz}, the eq.(\ref{sor}) allows a
cohomological interpretation of the cosmological constant, of the
Ashtekar Lagrangians \cite{ash} for pure gravity, of the
Capovilla, Jacobson, Dell Lagrangian \cite{cjd}
as well as the gravitational Chern-Simons terms.

\section{Maurer-Cartan horizontality conditions}
\setcounter{equation}{0}

The aim of this section is to derive the set of BRST
transformations for the Yang-Mills gauge fields in the presence
of gravity from Maurer-Cartan horizontality conditions.
In a first step this geometrical formalism is used to discuss
the simpler case of non-abelian Yang-Mills theory \cite{ba1}.

The BRST transformations of the 1-form gauge connection
$\AA^{A}=\AA^{A}_{\mu}dx^{\mu}$ and the 0-form
ghost field $c^{A}$ are given by\footnote{Here,
capital Latin indices are denoting gauge group indices.}
\bea
s\AA^A\=dc^{A}+f^{ABC}c^{B}\AA^{C} \ , \nonumber\\
sc^A\=\f1{2}f^{ABC}c^{B}c^{C} \ ,
\label{BRST_YM}
\eea
with
\eq
s^{2}=0 \ ,
\eqn{NILPOTENCY_YM}
where $f^{ABC}$ are the structure constants of the corresponding
gauge group ${\bf G}$.
As usual, the adopted grading is given by the sum of the
form degree and of the ghost number. In this sense, the fields
$\AA^{A}$ and $c^{A}$ are both of degree one,
their ghost number being respectively zero and one.
A $p$-form with ghost number $q$ will be denoted by
$\O^{q}_{p}$, its total grading being $(p+q)$.
The 2-form field strength $\FF^{A}$ is given by
\eq
\FF^{A}=\frac{1}{2}\FF^{A}_{\mu\nu}dx^{\mu}dx^{\nu}
=d\AA^{A}+\frac{1}{2}f^{ABC}\AA^{B}\AA^{C} \ ,
\eqn{FIELDSTRENGTH_YM}
and
\eq
d\FF^{A}=f^{ABC}\FF^{B}\AA^{C} \ ,
\eqn{BIANCHI_YM}
is its Bianchi identity.
The BRST transformations \equ{BRST_YM} can be interpreted as a
Maurer-Cartan horizontality condition if one introduces the combined
gauge-ghost field \cite{sto}
\eq
\wti{\AA}^{A}=\AA^{A}+c^{A} \ ,
\eqn{A_TILDE_YM}
and the generalized nilpotent differential operator
\eq
\wti{d}=d-s~~~,~~~\wti{d}^{2}=0 \ .
\eqn{d_TILDE_YM}
Notice that both $\wti{\AA}^{A}$ and $\wti{d}$ have degree one.
The nilpotency of $\wti{d}$ in \equ{d_TILDE_YM} just implies the
nilpotency of $s$ and $d$, and furthermore fulfills the
anticommutator relation
\eq
\lac s,d \rac = 0 \ .
\eqn{ANTICOMMUTATOR_YM}

With $\wti{d}$ and $\wti{\AA}$ we can build up the generalized
degree-two field strength $\wti{\FF}^{A}$:
\eq
\wti{\FF}^{A}=\wti{d}\wti{\AA}^{A}
+\frac{1}{2}f^{ABC}\wti{\AA}^{B}\wti{\AA}^{C} \ ,
\eqn{F_TILDE_YM}
which, from eq.\equ{d_TILDE_YM}, obeys the generalized Bianchi
identity
\eq
\wti{d}\wti{\FF}^{A}=f^{ABC}\wti{\FF}^{B}\wti{\AA}^{C} \ .
\eqn{GENERAL_BIANCHI_YM}
The Maurer-Cartan horizontality condition
reads then
\eq
\wti{\FF}^{A}=\FF^{A} \ .
\eqn{MCHC_YM}
Now it is very easy to check that the BRST transformations
\equ{BRST_YM} can be obtained from the horizontality condition
\equ{MCHC_YM} by simply expanding $\wti{\FF}^{A}$ in terms of the
elementary fields $\AA^{A}$ and $c^{A}$ and collecting the
terms with the same form degree and ghost number.

\subsection{Yang-Mills gauge fields coupled to gravity in Ashtekar
            variables}

If we want to study the BRST cohomology of the Yang-Mills fields in
the presence of gravity we should generalize the horizontality
condition \equ{MCHC_YM} and specify the functional space
the BRST operator $s$ acts upon. The latter is chosen to be the
space of local polynomials which depend on the 1-forms
$(e^{a},A^{a}_{~b},\AA^{A})$, where $e^{a}$, $A^{a}_{~b}$ and
$\AA^{A}$ being respectively the vielbein, the
Ashtekar connection and the Yang-Mills gauge field
\eqa
e^{a} \= e^{a}_{\mu}dx^{\mu} \ , \non
A^{a}_{~b} \= A^{a}_{~b\mu}dx^{\mu} \ , \non
\AA^{A} \= \AA^{A}_{\mu}dx^{\mu} \ ,
\eqan{OF}
and on the $2$-forms $(T^{a},F^{a}_{~b},\FF^{A})$,
whereby $T^{a}$, $F^{a}_{~b}$ and $\FF^{A}$
denoting the Ashtekar torsion, the Ashtekar
field strength and the Yang-Mills field strength
\eqa
T^{a} \= \frac{1}{2}T^{a}_{\mu\nu}dx^{\mu}dx^{\nu}
=de^{a}+A^{a}_{~b}e^{b}=De^{a} \ , \non
F^{a}_{~b} \= \frac{1}{2}F^{a}_{~b\mu\nu}dx^{\mu}dx^{\nu}
=dA^{a}_{~b}+A^{a}_{~c}A^{c}_{~b} \ , \non
\FF^{A} \= \frac{1}{2}\FF^{A}_{\mu\nu}dx^{\mu}dx^{\nu}
=d\AA^{A}+\frac{1}{2}f^{ABC}\AA^{B}\AA^{C} \ ,
\eqan{TF}
with the covariant exterior derivative
\eq
D=d+A+\AA \ .
\eqn{COVD}
The tangent space indices $(a,b,c,...)$ are referred to the
group $SO(1,3)$.

Applying the exterior derivative $d$ to both sides of eqs.\equ{TF}
one gets the Bianchi identities
\eqa
DT^{a} \= dT^{a}+A^{a}_{~b}T^{b}=F^{a}_{~b}e^{b} \ , \non
DF^{a}_{~b} \= dF^{a}_{~b}+A^{a}_{~c}F^{c}_{~b}
-A^{c}_{~b}F^{a}_{~c}=0 \ , \non
D\FF^{A} \= d\FF^{A}+f^{ABC}\AA^{B}\FF^{C}=0 \ .
\eqan{BI}

To write down the gravitational Maurer-Cartan horizontality
conditions for this case one introduces a further ghost,
as done in~\cite{tm1,ba1,tm2},
the local translation ghost $\h^{a}$ having ghost number one
and a tangent space index.
As explained in~\cite{tm1,tm2} (see also \cite{mss,mss1}),
the field  $\h^{a}$ represents the ghost of local translations
in the tangent space.

The local translation ghost $\h^{a}$ can be related
to the ghost of local diffeomorphisms $\x^{\mu}$ by
\eq
\x^{\mu}=E^{\mu}_{a}\h^{a}~~~,~~~\h^{a}=\x^{\mu}e^{a}_{\mu} \ ,
\eqn{ETA}
where $E^{\mu}_{a}$ denotes the inverse of the vielbein
$e^{a}_{\mu}$, i.e.
\eqa
e^{a}_{\mu}E^{\mu}_{b} \=~ \d^{a}_{b} \ ,\non
e^{a}_{\mu}E^{\nu}_{a} \=~ \d^{\nu}_{\mu} \ .
\eqan{VIEL_ORTHO}
Proceeding now as for the pure Yang-Mills case, one defines the
nilpotent differential operator $\wti{d}$ of degree one
\eq
\wti{d}=d-s \ ,
\eqn{EXTD}
and the generalized vielbein-ghost field $\ti{e}^{a}$, the extended
Ashtekar connection $\wti{A}^{a}_{~b}$, and the generalized
non-abelian Yang-Mills gauge field $\wti{\AA}^A$
\eqa
\ti{e}^{a} \= e^{a}+\h^{a} \ , \non
\wti{A}^{a}_{~b} \= \widehat{A}^{a}_{~b}+c^{a}_{~b} \ , \non
\wti{\AA}^{A} \= \widehat{\AA}^{A}+c^{A} \ ,
\eqan{EVIEL}
with the ashtekar ghost $c^{a}_{~b}$ and
where $\widehat{A}^{a}_{~b}$ and $\widehat{\AA}^{A}$ are given by
\eqa
\widehat{A}^{a}_{~b}\=A^{a}_{~bm}\ti{e}^{m}
=A^{a}_{~b}+A^{a}_{~bm}\h^{m} \ , \non
\widehat{\AA}^{A}\=\AA^{A}_{m}\ti{e}^{m}
=\AA^{A}+\AA^{A}_{m}\h^{m} \ ,
\eqan{HCO}
with the 0-forms $A^{a}_{~bm}$\footnote{Remark that the
0-form $A^{a}_{~bm}$ does not possess any symmetric or
antisymmetric property with respect to the lower indices $(bm)$.}
and $\AA^{A}_{m}$ defined by the expansion of the 0-form
Ashtekar connection $A^{a}_{~b\mu}$ and the 0-form Yang-Mills gauge
field $\AA^{A}_{\m}$ in terms of the vielbein $e^{a}_{\mu}$,
i.e.:
\eqa
A^{a}_{~b\mu}\= A^{a}_{~bm}e^{m}_{\mu} \ , \non
\AA^{A}_{\mu}\= \AA^{A}_{m}e^{m}_{\mu} \ .
\eqan{SPINC}
As it is well-known, the last formulas stem from the fact that the
vielbein formalism allows to transform locally the spacetime
indices of an arbitrary
tensor ${\cal N}_{\mu\nu\rho\sigma...}$ into flat tangent space
indices $\NN_{abcd...}$ by means of the expansion
\eq
\NN_{\mu\nu\rho\sigma...}=\NN_{abcd...}
e^{a}_{\mu}e^{b}_{\nu}e^{c}_{\rho}e^{d}_{\sigma}...  \ .
\eqn{WORLD}
Vice versa one has
\eq
\NN_{abcd...}=\NN_{\mu\nu\rho\sigma...}
E^{\mu}_{a}E^{\nu}_{b}E^{\rho}_{c}E^{\sigma}_{d}...  \ .
\eqn{TANG}
According to eqs.\equ{TF}, the generalized Ashtekar torsion field,
the generalized Ashtekar field strength and the generalized
Yang-Mills field strength are given by
\eqa
\wti{T}^{a} \= \wti{d}\ti{e}^{a}+\wti{A}^{a}_{~b}\ti{e}^{b}
=\wti{D}\ti{e}^{a} \ , \non
\wti{F}^{a}_{~b} \= \wti{d}\wti{A}^{a}_{~b}+\wti{A}^{a}_{~c}
\wti{A}^{c}_{~b} \ , \non
\wti{\FF}^{A}\=\wti{d}\wti{\AA}^{A}
+\frac{1}{2}f^{ABC}\wti{\AA}^{B}\wti{\AA}^{C} \ ,
\eqan{DEFTWO}
and are easily seen to obey the generalized Bianchi identities
\eqa
\wti{D}\wti{T}^{a} \= \wti{d}\wti{T}^{a}
+\wti{A}^{a}_{~b}\wti{T}^{b}
=\wti{F}^{a}_{~b}\ti{e}^{b} \ , \non
\wti{D}\wti{F}^{a}_{~b} \= \wti{d}\wti{F}^{a}_{~b}
+\wti{A}^{a}_{~c}\wti{F}^{c}_{~b}
-\wti{A}^{c}_{~b}\wti{F}^{a}_{~c}=0 \ , \non
\wti{D}\wti{\FF}^{A}\=\wti{d}\wti{\FF}^{A}
+f^{ABC}\wti{\AA}^{B}\wti{\FF}^{C}=0 \ ,
\eqan{GBI}
with
\eq
\wti{D}=\wti{d}+\wti{A}+\wti{\AA}
\eqn{GENERAL_COVD}
the generalized covariant derivative.

With these definitions the Maurer-Cartan horizontality conditions
for the Yang-Mills gauge fields in the presence of gravity in terms
of Ashtekar fields may be expressed in the following way:
{\it $\ti{e}$ and all its generalized covariant
exterior differentials can be expanded over $\ti{e}$ with
classical coefficients},
\eqa
\label{MCG1}
\ti{e}^{a}\=\d^{a}_{b}\ti{e}^{b} \equiv horizontal \ , \\
\label{MCG2}
\wti{T}^{a}(\ti{e},\wti{A})
\=\f1{2}T^{a}_{mn}(e,A)\ti{e}^{m}\ti{e}^{n}
\equiv horizontal \ , \\
\label{MCG3}
\wti{F}^{a}_{~b}(\wti{A})
\=\frac{1}{2}F^{a}_{~bmn}(A)\ti{e}^{m}\ti{e}^{n}
\equiv horizontal \ , \\
\label{MCG4}
\wti{\FF}^{A}(\wti{\AA})
\=\frac{1}{2}\FF^{A}_{mn}(\AA)\ti{e}^{m}\ti{e}^{n}
\equiv horizontal \ .
\eqan{MAURER_CARTAN_HC}
Through eq.\equ{WORLD}, the 0-forms $T^{a}_{mn}$,
$F^{a}_{~bmn}$, and $\FF^{A}_{mn}$
are defined by the vielbein expansion of the 2-forms of the
Ashtekar torsion, the Ashtekar field strength and the
Yang-Mills field strength of eqs.\equ{TF},
\eqa
T^{a}\=\frac{1}{2}T^{a}_{mn}e^{m}e^{n} \ , \non
F^{a}_{~b}\=\frac{1}{2}F^{a}_{~bmn}e^{m}e^{n} \ , \non
\FF^{A}\=\frac{1}{2}\FF^{A}_{mn}e^{m}e^{n} \ ,
\eqan{TWOFORM}
and the 0-form $D_{m}$ of the covariant exterior derivative $D$
is given by
\eq
D=e^{m}D_{m} \ .
\eqn{ED}
Notice also that eqs.\equ{HCO} are nothing but the horizontality
conditions for the Ashtekar connection and the Yang-Mills gauge
field expressing the fact that
$\widehat{A}$ and $\widehat{\AA}$
themselves can be expanded over $\ti{e}$.

The horizontality conditions (\ref{MCG1})-(\ref{MCG4}) are
equivalent with the statements
\bea
\tilde{e}^a\=exp(i_{\xi})e^a=e^a+i_{\xi}e^a \ , \\
\wti{T}^a\=exp(i_{\xi})T^a=T^a+i_{\xi}T^a
+\frac{1}{2}i_{\xi}i_{\xi}T^a \ , \\
\wti{F}^{ab}\=exp(i_{\xi})F^{ab}=F^{ab}+i_{\xi}F^{ab}
+\frac{1}{2}i_{\xi}i_{\xi}F^{ab} \ , \\
\wti{\FF}^A\=exp(i_{\xi})\FF^A=\FF^A+i_{\xi}\FF^A
+\frac{1}{2}i_{\xi}i_{\xi}\FF^A \ ,
\eea
since $e^a$ is a 1-form, while $T^a$, $F^{ab}$ and $\FF^A$
are 2-forms.

Eqs.\equ{MCG1}-\equ{MCG4} define the Maurer-Cartan horizontality
conditions for the Yang-Mills gauge fields in the presence of
gravity in terms of Ashtekar variables and when expanded in
terms of the elementary fields
$(e^{a}, A^{a}_{~b}, \AA^{A}, \h^{a}, c^{a}_{~b}, c^{A})$,
give the nilpotent BRST transformations corresponding to the
diffeomorphism transformations, the local Lorentz rotations and
the gauge transformations.

For a better understanding of this point let us discuss in detail
the horizontality condition \equ{MCG2} for the Ashtekar torsion.
Making use of eqs.\equ{EXTD}, \equ{EVIEL}, \equ{HCO} and of the
definition \equ{DEFTWO}, one verifies that eq.\equ{MCG2} gives
\eqa
de^{a}-se^{a}+d\h^{a}-s\h^{a}+A^{a}_{~b}e^{b}
+c^{a}_{~b}e^{b}
+A^{a}_{~b}\h^{b}+c^{a}_{~b}\h^{b}
+A^{a}_{~bm}\h^{m}e^{b}+A^{a}_{~bm}\h^{m}\h^{b}=&&\non
=\frac{1}{2}T^{a}_{mn}e^{m}e^{n}+T^{a}_{mn}e^{m}\h^{n}
+\frac{1}{2}T^{a}_{mn}\h^{m}\h^{n} \ ,~~~&&
\eqan{EXPANSION}
from which, collecting the terms with the same form degree and
ghost number, one easily obtains the BRST transformations for the
tetrad 1-form $e^{a}$ and for the local translation ghost $\h^{a}$:
\eqa
se^{a}\=d\h^{a}+A^{a}_{~b}\h^{b}+c^{a}_{~b}e^{b}
+A^{a}_{~bm}\h^{m}e^{b}-T^{a}_{mn}e^{m}\h^{n} \ ,\non
s\h^{a}\=c^{a}_{~b}\h^{b}+A^{a}_{~bm}\h^{m}\h^{b}
-\frac{1}{2}T^{a}_{mn}\h^{m}\h^{n} \ .
\eqan{BRSE}
These equations, when rewritten in terms of the variable $\x^{\mu}$
of eq.\equ{ETA}, take the more familiar form
\eqa
se^{a}_{\mu}\=c^{a}_{~b}e^{b}_{\mu}-\LL_{\x}e^{a}_{\mu} \ , \non
s\x^{\mu}\=-\x^{\l}\6_{\l}\x^{\mu}=-\frac{1}{2}\LL_{\x}\x^{\mu} \ ,
\eqan{BRS_WORLD}
where $\LL_{\x}$ denotes the ordinary Lie derivative along the
direction $\x^{\mu}$, i.e.
\eq
\LL_{\x}e^{a}_{\mu}=\x^{\l}\6_{\l}e^{a}_{\mu}
+(\6_{\mu}\x^{\l})e^{a}_{\l} \ .
\eqn{LIE_DERIV}
It is apparent now that eq.\equ{BRSE} represents the tangent space
formulation of the usual BRST transformations corresponding to local
Lorentz rotations and diffeomorphisms.

One sees then that the Maurer-Cartan horizontality conditions
\equ{MCG1}-\equ{MCG4} together with eq.\equ{DEFTWO} carry in a very
simple and compact way all the information relative to the
gravitational Yang-Mills gauge algebra. Indeed, it is easy to expand
eqs.\equ{MCG1}-\equ{MCG4} in terms of $e^{a}$ and $\h^{a}$ and work
out the BRST transformations of the remaining fields
$(A^{a}_{~b},\AA^{A},T^{a},F^{a}_{~b},\FF^{A},c^{a}_{~b},c^{A})$.

However, in view of the fact that we will use as fundamental
variables the 0-forms $(A^{a}_{~bm}, \AA^{A}_{m}, T^{a}_{mn},
F^{a}_{~bmn}, \FF^{A}_{mn})$ rather than
the 1-forms $A^{a}_{~b}$ and $\AA^{A}$ and the 2-forms
$T^{a}$, $F^{a}_{~b}$ and $\FF^{A}$ let us proceed by introducing
the partial derivative $\6_{a}$ with a flat tangent space index.
According to the formulas \equ{WORLD} and \equ{TANG}, the latter is
defined by
\eq
\6_{a} \equiv E^{\mu}_{a}\6_{\mu} \ ,
\eqn{TANGENT_DERIV}
and
\eq
\6_{\mu} = e^{a}_{\mu}\6_{a} \ ,
\eqn{WORLD_DERIV}
so that the intrinsic exterior differential $d$ becomes
\eq
d=dx^{\mu}\6_{\mu}=e^{a}\6_{a} \ .
\eqn{EXTER_DERIV}
Let us emphasize that the introduction of the operator $\6_{a}$ and
the use of the 0-forms $(A^{a}_{~bm}, \AA^{A}_{m}, T^{a}_{mn},
F^{a}_{~bmn}, \FF^{A}_{mn})$
allows for a complete tangent space formulation of the gravitational
Yang-Mills gauge algebra.
This step, as we shall see later, turns out to be very
useful in the analysis of the corresponding BRST cohomology.
Moreover, as one can easily understand, the knowledge of the BRST
transformations
of the 0-form sector $(A^{a}_{~bm}, \AA^{A}_{m}, T^{a}_{mn},
F^{a}_{~bmn}, \FF^{A}_{mn})$
together with the expansions \equ{SPINC}, \equ{TWOFORM} and the
eqs.\equ{BRSE} completely characterize the transformation
law of the forms $(A^{a}_{~b},\AA^{A},T^{a},F^{a}_{~b},\FF^{A})$.

\subsection{BRST transformations and Bianchi identities}

Let us finish this section by giving, for the convenience of the
reader, the BRST transformations and the Bianchi identities which
one can find by using the Maurer-Cartan horizontality conditions
\equ{MCG1}-\equ{MCG4} and from eqs.\equ{DEFTWO} and \equ{GBI} for
each form sector and ghost number.

\begin{itemize}

\item {\bf Form sector two, ghost number zero
  $(T^{a}, F^{a}_{~b}, \FF^{A})$}

\eqa
sT^{a}\=c^{a}_{~b}T^{b}+A^{a}_{~bk}\h^{k}T^{b}
-F^{a}_{~b}\h^{b}\non
\+A^{a}_{~b}T^{b}_{mn}e^{m}\h^{n}-F^{a}_{~bmn}e^{b}e^{m}\h^{n}
+(dT^{a}_{mn})e^{m}\h^{n}\non
\-T^{a}_{mn}e^{m}d\h^{n}+T^{a}_{mn}T^{m}\h^{n}
-T^{a}_{kn}A^{k}_{~m}e^{m}\h^{n} \ , \non
sF^{a}_{~b}\=c^{a}_{~c}F^{c}_{~b}-c^{c}_{~b}F^{a}_{~c}
+A^{a}_{~ck}\h^{k}F^{c}_{~b}-A^{c}_{~bk}\h^{k}F^{a}_{~c}\non
\+A^{a}_{~c}F^{c}_{~bmn}e^{m}\h^{n}
-A^{c}_{~b}F^{a}_{~cmn}e^{m}\h^{n}+(dF^{a}_{~bmn})e^{m}\h^{n}\non
\+F^{a}_{~bmn}T^{m}\h^{n}
-F^{a}_{~bkn}A^{k}_{~m}e^{m}\h^{n}
-F^{a}_{~bmn}e^{m}d\h^{n} \ , \non
s\FF^{A}\=(d\FF^{A}_{mn})e^{m}\h^{n}+\FF^{A}_{mn}T^{m}\h^{n}
-\FF^{A}_{mn}A^{m}_{~k}e^{k}\h^{n}-\FF^{A}_{mn}e^{m}d\h^{n}\non
\+f^{ABC}c^{B}\FF^{C}+f^{ABC}\AA^{B}_{m}\h^{m}\FF^{C}
+f^{ABC}\AA^{B}\FF^{C}_{mn}e^{m}\h^{n} \ .
\eqan{FORMTWO}
For the Bianchi identities one has
\eqa
&&dT^{a}+A^{a}_{~b}T^{b}=F^{a}_{~b}e^{b} \ , \non
&&dF^{a}_{~b}+A^{a}_{~c}F^{c}_{~b}
-A^{c}_{~b}F^{a}_{~c}=0 \ , \non
&&d\FF^{A}+f^{ABC}\AA^{B}\FF^{C}=0 \ .
\eqan{BIG}

\item {\bf Form sector one, ghost number zero
 $(e^{a}, A^{a}_{~b}, \AA^{A})$}

\eqa
se^{a}\=d\h^{a}+A^{a}_{~b}\h^{b}+c^{a}_{~b}e^{b}
+A^{a}_{~bm}\h^{m}e^{b}
-T^{a}_{mn}e^{m}\h^{n} \ , \non
sA^{a}_{~b}\=dc^{a}_{~b}+c^{a}_{~c}A^{c}_{~b}
+A^{a}_{~c}c^{c}_{~b}
+(dA^{a}_{~bm})\h^{m}+A^{a}_{~bm}d\h^{m}\non
\+A^{a}_{~c}A^{c}_{~bm}\h^{m}+A^{a}_{~cm}\h^{m}A^{c}_{~b}
-F^{a}_{~bmn}e^{m}\h^{n} \ , \non
s\AA^{A}\=dc^{A}+(d\AA^{A}_{m})\h^{m}+\AA^{A}_{m}d\h^{m}
+f^{ABC}\AA^{B}c^{C}\non
\+f^{ABC}\AA^{B}\AA^{C}_{m}\h^{m}
-\FF^{A}_{mn}e^{m}\h^{n} \ .
\eqan{FORM1}


\item {\bf Form sector zero, ghost number zero
$(A^{a}_{~bm}, \AA^{A}_{m}, T^{a}_{mn}, F^{a}_{~bmn},
\FF^{A}_{mn})$ }

\eqa
sA^{a}_{~bm}\=-\6_{m}c^{a}_{~b}+c^{a}_{~c}A^{c}_{~bm}
-c^{c}_{~b}A^{a}_{~cm}
-c^{k}_{~m}A^{a}_{~bk}-\h^{k}\6_{k}A^{a}_{~bm} \ , \non
s\AA^{A}_{m}\=-\6_{m}c^{A}-f^{ABC}\AA^{B}_{m}c^{C}
-c^{k}_{~m}\AA^{A}_{k}-\h^{k}\6_{k}\AA^{A}_{m} \ , \non
sT^{a}_{mn}\=c^{a}_{~k}T^{k}_{mn}-c^{k}_{~m}T^{a}_{kn}
-c^{k}_{~n}T^{a}_{mk}-\h^{k}\6_{k}T^{a}_{mn} \ , \non
sF^{a}_{~bmn}\=c^{a}_{~c}F^{c}_{~bmn}-c^{c}_{~b}F^{a}_{~cmn}
-c^{k}_{~m}F^{a}_{~bkn}-c^{k}_{~n}F^{a}_{~bmk}
-\h^{k}\6_{k}F^{a}_{~bmn} \ , \non
s\FF^{A}_{mn}\=f^{ABC}c^{B}\FF^{C}_{mn}-c^{k}_{~m}\FF^{A}_{kn}
-c^{k}_{~n}\FF^{A}_{mk}-\h^{k}\6_{k}\FF^{A}_{mn} \ .
\eqan{0_FORMS}
The Bianchi identities \equ{BIG} are projected on the 0-form
Ashtekar torsion $T^{a}_{mn}$, on the 0-form Ashtekar field
strength $F^{a}_{~bmn}$
and on the 0-form Yang-Mills field strength $\FF^{A}_{mn}$
to give
\bea
dT^{a}_{mn}\=(\6_{l}T^{a}_{mn})e^{l}\nonumber\\
\=(F^{a}_{~lmn}+F^{a}_{~mnl}+F^{a}_{~nlm}\nonumber\\
\-A^{a}_{~bl}T^{b}_{mn}-A^{a}_{~bm}T^{b}_{nl}
-A^{a}_{~bn}T^{b}_{lm}\nonumber\\
\+T^{a}_{kn}T^{k}_{ml}+T^{a}_{km}T^{k}_{ln}
+T^{a}_{kl}T^{k}_{nm}\nonumber\\
\-T^{a}_{kn}A^{k}_{~lm}-T^{a}_{km}A^{k}_{~nl}
-T^{a}_{kl}A^{k}_{~mn}\nonumber\\
\+T^{a}_{kn}A^{k}_{~ml}+T^{a}_{kl}A^{k}_{~nm}
+T^{a}_{km}A^{k}_{~ln}\nonumber\\
\-\6_{m}T^{a}_{nl}-\6_{n}T^{a}_{lm})e^{l} \ , \nonumber\\
dF^{a}_{~bmn}\=(\6_{l}F^{a}_{~bmn})e^{l}\nonumber\\
\=(-A^{a}_{~cl}F^{c}_{~bmn}-A^{a}_{~cm}F^{c}_{~bnl}
-A^{a}_{~cn}F^{c}_{~blm}\nonumber\\
\+A^{c}_{~bl}F^{a}_{~cmn}+A^{c}_{~bm}F^{a}_{~cnl}
+A^{c}_{~bn}F^{a}_{~clm}\nonumber\\
\+F^{a}_{~bkn}T^{k}_{ml}+F^{a}_{~bkm}T^{k}_{ln}
+F^{a}_{~bkl}T^{k}_{nm}\nonumber\\
\-F^{a}_{~bkn}A^{k}_{~lm}-F^{a}_{~bkm}A^{k}_{~nl}
-F^{a}_{~bkl}A^{k}_{~mn}\nonumber\\
\+F^{a}_{~bkn}A^{k}_{~ml}+F^{a}_{~bkl}A^{k}_{~nm}
+F^{a}_{~bkm}A^{k}_{~ln}\nonumber\\
\-\6_{m}F^{a}_{~bnl}-\6_{n}F^{a}_{~blm})e^{l} \ , \nonumber\\
d\FF^{A}_{mn}\=(\6_{l}\FF^{A}_{mn})e^{l}\nonumber\\
\=(f^{ABC}\FF^{B}_{mn}\AA^{C}_{l}+f^{ABC}\FF^{B}_{nl}\AA^{C}_{m}
+f^{ABC}\FF^{B}_{lm}\AA^{C}_{n}\nonumber\\
\-\FF^{A}_{kn}T^{k}_{lm}-\FF^{A}_{kl}T^{k}_{mn}
-\FF^{A}_{km}T^{k}_{nl}\nonumber\\
\+\FF^{A}_{kn}A^{k}_{~ml}+\FF^{A}_{km}A^{k}_{~ln}
+\FF^{A}_{kl}A^{k}_{~nm}\nonumber\\
\-\FF^{A}_{kn}A^{k}_{~lm}-\FF^{A}_{kl}A^{k}_{~mn}
-\FF^{A}_{km}A^{k}_{~nl}\nonumber\\
\-\6_{m}\FF^{A}_{nl}-\6_{n}\FF^{A}_{lm})e^{l} \ .
\label{0_BI}
\eea
One has also the equations
\eqa
dA^{a}_{~bm}\=(\6_{n}A^{a}_{~bm})e^{n}\non
\=(-F^{a}_{~bmn}+A^{a}_{~cm}A^{c}_{~bn}
-A^{a}_{~cn}A^{c}_{~bm}\non
\+A^{a}_{~bk}T^{k}_{mn}-A^{a}_{~bk}A^{k}_{~nm}
+A^{a}_{~bk}A^{k}_{~mn}+\6_{m}A^{a}_{~bn})e^{n} \ , \non
d\AA^{A}_{m}\=(\6_{n}\AA^{A}_{m})e^{n}\non
\=(-\FF^{A}_{mn}+f^{ABC}\AA^{B}_{m}\AA^{C}_{n}
+\AA^{A}_{k}T^{k}_{mn}\non
\-\AA^{A}_{k}A^{k}_{~nm}+\AA^{A}_{k}A^{k}_{~mn}
+\6_{m}\AA^{A}_{n})e^{n} \ .
\eqan{0_DEF}

\item {\bf Form sector zero, ghost number one
$(\h^{a}, c^{a}_{~b}, c^{A})$ }

\eqa
s\h^{a}\=A^{a}_{~bm}\h^{m}\h^{b}+c^{a}_{~b}\h^{b}
-\frac{1}{2}T^{a}_{mn}\h^{m}\h^{n} \ , \non
sc^{a}_{~b}\=c^{a}_{~c}c^{c}_{~b}
-\h^{k}\6_{k}c^{a}_{~b} \ , \non
sc^{A}\=\frac{1}{2}f^{ABC}c^{B}c^{C}-\h^{k}\6_{k}c^{A} \ .
\eqan{FORMZERO}

\item {\bf Algebra between $s$ and $d$ }

{}From the above transformations it follows:
\eq
s^{2}=0~~~,~~~d^{2}=0 \ ,
\eqn{S_D_ALGEBRA_1}
and
\eq
\{s,d\}=0 \ .
\eqn{S_D_ALGEBRA_2}

\end{itemize}

\section{Solution of the descent equations}
\setcounter{equation}{0}

The question of finding the invariant Lagrangians,
the anomalies and the Schwinger terms for the
Yang-Mills gauge field theory coupled to four-dimensional gravity in
Ashtekar variables can be solved in a purely algebraic way by
solving the BRST consistency condition in the space of the
integrated local field polynomials.
In order to solve this problem we have to find out the nontrivial
solution of the Wess-Zumino consistency condition \cite{wz}
\be
s\Delta =0 \label{e} \ ,
\ee
where $\Delta $ is an integrated local field polynomial, i.e.
$\Delta =\int{\cal B}$.
The condition~(\ref{e}) translates into the local equation
\be
s{\cal B}+d{\cal Q}=0 \label{sa} \ ,
\ee
where ${\cal Q}$ is some local polynomial and
$d=dx^\mu \partial _\mu $ is the nilpotent exterior spacetime
derivative which anticommutes with
the nilpotent BRST operator $s$
\be
s^2=d^2=sd+ds=0 \label{sd}
\ee
and it is {\em acyclic} (i.e. its cohomology group vanishes).

The local equation~(\ref{sa}), due to~(\ref{sd}) and the
acyclicity of $d$, generates a tower of descent equations
\bea
s{\cal B}+d{\cal Q}^1=0 \nonumber\\
s{\cal Q}^1+d{\cal Q}^2=0 \nonumber\\
\cdots \nonumber\\ s{\cal Q}^{k-1}+d{\cal Q}^k=0 \nonumber\\
s{\cal Q}^k=0
\label{des.eq}
\eea
with ${\cal Q}^i$ local polynomials in the fields.

For the Yang-Mills case, these equations can be solved by means of a
transgression procedure generated by the {\em Russian formula}
\cite{tm1,sto,witten,ba1,dvio,gins,ton1,stora,brandt}.

More recently a new and efficient way of finding nontrivial
solutions of the tower (\ref{des.eq}) has been proposed by
S.P. Sorella \cite{sor} and successfully applied to the study of the
Yang-Mills cohomology \cite{st}, the gravitational anomalies
\cite{ws} and the algebraic structure of gravity with torsion
\cite{mss}. The basic ingredient of the method is an operator
$\delta$ which allows us to express the exterior derivative $d$ as
a BRST commutator, i.e.:
\be
d=-[s,\delta] \ .
\label{comm}
\ee

Now it is easy to see that, once the decomposition (\ref{comm}) has
been found, repeated application of the operator $\delta $ on the
polynomial ${\cal Q}$ which is a nontrivial solution of the last
equation of (\ref{des.eq}) gives an explicit and nontrivial
solution for the other cocycles ${\cal Q}^i$ and for
${\cal B}$.
If ${\cal B}$ has ghost number one then it is called an anomaly and
if it has ghost number zero then it represents an invariant
Lagrangian. In other word using the operator $\delta$ we can
calculate the solution of the cohomology $H(s~mod~d)$ if we
know the solution of the cohomology $H(s)$.
Actually, as
has been shown in \cite{st}, the cocycles obtained by the descent
equations~(\ref{des.eq}) turn out to be completely equivalent to
those which are based on the {\em Russian formula}.

For the Yang-Mills fields coupled with gravity in the Ashtekar
variables the operator $\delta$ introduced in (\ref{comm}) can be
defined by
\bea
\delta\eta^a \= -e^a \ , \nonumber\\
\delta \Phi \= 0 \hspace{0.8cm}
\mbox{for}
\hspace{0.8cm}
\Phi=(e^a,A^{ab},\AA^A,T^a,F^{ab},\FF^A,c^{ab},c^A) \ .
\label{delta}
\eea
Now it is easy to verify that $\delta $ is of degree 0 and obeys the
following algebraic relations
\begin{equation}
\label{alg}d=- [s,\delta ] \hspace{0.7cm},
\hspace{0.7cm}[d,\delta]=0 \ .
\end{equation}
In order to solve the tower (\ref{des.eq}) we shall make use of the
following identity
\be
e^{\delta}s=(s+d)e^{\delta} \ ,
\label{id}
\ee
which is a direct consequence of (\ref{alg}) (see \cite{st}).

Let us consider now the solution of eqs.(\ref{des.eq}) with a given
ghost number $G$ and form degree $4$, i.e. a solution of the tower
\bea
s\Omega _4^G+d\Omega _3^{G+1}=0 \nonumber\\
s\Omega _3^{G+1}+d\Omega_2^{G+2}=0 \nonumber\\
s\Omega _2^{G+2}+d\Omega _1^{G+3}=0 \nonumber\\
s\Omega _1^{G+3}+d\Omega _0^{G+4}=0 \nonumber\\
s\Omega _0^{G+4}=0
\label{des1.eq}
\eea
with ($\Omega _4^G$, $\Omega _3^{G+1}$, $\Omega _2^{G+2}$,
$\Omega _1^{G+3}$, $\Omega_0^{G+4}$) local polynomials in the
variables
($e^a$, $A^{ab}$, $\AA^A$, $\eta^a$, $c^{ab}$, $c^A$)
which, without loss of generality, will be always considered as
irreducible elements,
i.e. they cannot be expressed as the product of
several factored terms. In particular $\Omega _4^0,\Omega _3^1$ and
$\Omega _2^2$ correspond, respectively to an invariant Lagrangian,
an anomaly and a Schwinger term.

Due to the identity (\ref{id}) we can obtain the higher cocycles
$\Omega^{G+4-q}_q (q=1,2,3,4)$ once a nontrivial solution for
$\Omega^{G+4}_0$ is known. Indeed, if one applies the
identity (\ref{id}) on $\Omega^{G+4}_0$ one gets
\be
(s+d)\left[e^\delta\Omega^{G+4}_0(\eta^a,c^{ab},c^{ab},c^{A},A^{ab},
\AA^A,T^a,F^{ab},\FF^A)\right]=0 \ .
\label{my}
\ee
But as one can see from eq.(\ref{delta}), the operator $\delta$
acts as a shift on the ghost $\eta^a$ with an amount $(-e^a)$
and eq.(\ref{my}) can be rewritten as
\be
(s+d)\Omega^{G+4}_0(\eta-e,c,A,T,F)=0 \ .
\label{my1}
\ee
Thus the expansion of the 0-form cocycle
$\Omega^{G+4}_0(\eta^a-e^a,c^{ab},c^A,A^{ab},\AA^A,
\\T^a,F^{ab},\FF^A)$
in power of the 1-form tetrads $e^a$ yields all the cocycles
$\Omega^{G+4-q}_q$.

\section{Examples}
\setcounter{equation}{0}

This section is devoted to apply the previous algebraic setup and
to discuss some explicit examples.
We want to emphasize the cohomological
origin of the Lagrangian which describes Yang-Mills fields in the
presence of gravity, the topological Yang-Mills Lagrangian
as well as of the Chern-Simons terms for this theory.

\subsection{The Yang-Mills Lagrangian in the presence of gravity}

The simplest  local BRST polynomial which one can construct
from the Yang-Mills fields and the local translation ghost is
\be
\Omega_0^4=\f{1}{4}Tr(\FF^{mn}\FF_{mn})\f{1}{4!}
\varepsilon_{abcd}\eta^a\eta^b\eta^c\eta^d \ ,
\ee
with $\varepsilon_{abcd}$ the totally antisymmetric invariant
tensor of $SO(1,3)$.
Taking into account that in 4-dimensional spacetime
the product of 5 ghost fields $\eta^a$ automatically vanishes, it is
easy to check that $\Omega_0^4$ identifies a cohomology class
of the BRST differential, i.e.
\be
s\Omega_0^4=0~~~,~~~\Omega_0^4\neq s\widehat{\Omega}_0^3 \ .
\ee

The 0-form cocycle corresponds to the invariant Yang-Mills
Lagrangian in the presence of gravitational fields
\bea
\Omega_4^0\=\f{\delta^4}{4!}\Omega_0^4=\f1{4}Tr(\FF^{mn}\FF_{mn})
\f1{4!}\varepsilon_{abcd}e^ae^be^ce^d \nonumber\\
\=\f1{4}Tr(\FF^{mn}\FF_{mn})ed^4x=\f1{4}Tr(\FF^{\mu\nu}\FF_{mn})
e^m_{\mu}e^n_{nu}ed^4x \nonumber\\
\=\f1{4}Tr(\FF_{\mu\nu}\FF_{\tau\sigma})g^{\mu\tau}g^{\nu\sigma}
\sqrt{-g}d^4x \ ,
\eea
where $e=det(e^a_{\mu})=\sqrt{-g}$ denotes the determinant.

\subsection{The topological Yang-Mills Lagrangian}

We start with the local ghost polynomial
\be
\widehat{\Omega}^3_0=Tr\lp\widehat{\FF}\hat{c}-\f1{2}\hat{c}
\hat{c}\hat{c}\rp=Tr\lp(s\hat{c})\hat{c}-\f{2}{3}
\hat{c}\hat{c}\hat{c}\rp \ ,
\label{LGP}
\ee
where we have used the following redefined ghost variables
\be
\hat{c}^A=c^A+\AA^A_m\eta^m~~~,~~~
\widehat{\FF}^{A}=\frac{1}{2}\FF^{A}_{mn}\h^{m}\h^{n} \ ,
\label{RGB}
\ee
which have the BRST transformations
\bea
s\hat{c}^A\=\f1{2} f^{ABC}\hat{c}^B\hat{c}^C-
\widehat{\FF}^A \ , \nonumber\\
s\widehat{\FF}^A\=f^{ABC}\hat{c}^B\widehat{\FF}^C \ .
\eea
Under the action of the operator $\d$ they transform as
\be
\delta \hat{c}^A=-\AA^A~~~,~~~\delta\widehat{\FF}^A=-
\FF^A_{mn}e^m\eta^n \ .
\ee
Starting with the trivial cocycle
\bea
\Omega^4_0=s\widehat{\Omega}^3_0 \= Tr(\widehat{\FF}
\widehat{\FF})\nonumber\\
\=\f1{4}Tr(\FF_{kl}\FF_{mn})\eta^k\eta^l\eta^m\eta^n \ ,
\eea
we obtain the topological Yang-Mills invariant Lagrangian:
\bea
\Omega^0_4\=\f1{4!}\delta^4\Omega^4_0=Tr(\FF \FF)\nonumber\\
\=\f1{4}Tr(\FF_{kl}\FF_{mn})e^ke^le^me^n=\f1{4}Tr(\FF_{kl}\FF_{mn})
\varepsilon^{klmn} e d^4x \nonumber\\
\=\f1{4}Tr(\FF_{\mu\nu}\FF_{\tau\sigma})dx^{\mu}dx^{\nu}dx^{\tau}
dx^{\sigma}=\f1{4}Tr(\FF_{\mu\nu}\FF_{\tau\sigma})\varepsilon^{\mu
\nu\tau\sigma}d^4x \ .
\eea

\subsection{Chern-Simons terms, gauge anomalies and Ashtekar
            variables}

In our theory we can define two types of Chern-Simons Lagrangians:
one for the Yang-Mills gauge field $\AA^A$ and the other
for the Ashtekar connection $A^a_{~b}$.

For the sake of clarity and to make contact with the results
obtained in~\cite{sor,fb}, let us discuss in detail the construction
of the 3-dimensional Chern-Simons term.
In this case the descent equations take the form
\eqa
&&s\O^{0}_{3}+d\O^{1}_{2}=0 \ ,\non
&&s\O^{1}_{2}+d\O^{2}_{1}=0 \ ,\non
&&s\O^{2}_{1}+d\O^{3}_{0}=0 \ ,\non
&&s\O^{3}_{0}=0 \ ,
\eqan{TOWER_CS_3}
and thus their solutions are:
\eqa
&&\O^{2}_{1}=\d\O^{3}_{0} \ ,\non
&&\O^{1}_{2}=\frac{\d^{2}}{2!}\O^{3}_{0} \ ,\non
&&\O^{0}_{3}=\frac{\d^{3}}{3!}\O^{3}_{0} \ .
\eqan{HICO_3}
In order to find a solution for $\O^{3}_{0}$ we use again the
redefined ghost variables:
\bea
\hat{c}^{A}=\AA^{A}_{m}\h^{m}+c^{A}~~~,~~~
\widehat{\FF}^{A}=\frac{1}{2}\FF^{A}_{mn}\h^{m}\h^{n} \ .
\label{C_HAT}
\eea

For the cocycle $\O^{3}_{0}$ one then has
\eq
\O^{3}_{0}=\frac{1}{3!}f^{ABC}\hat{c}^{A}\hat{c}^{B}\hat{c}^{C}
-\widehat{\FF}^{A}\hat{c}^{A} \ ,
\eqn{ZEROFORM_CS_3}
from which $\O^{2}_{1}$, $\O^{1}_{2}$, and $\O^{0}_{3}$ are
computed to be
\bea
\O^{2}_{1}\=-\frac{1}{2}f^{ABC}\AA^{A}\hat{c}^{B}\hat{c}^{C}
+\widehat{\FF}^{A}\AA^{A}+\FF^{A}_{mn}e^{m}\h^{n}\hat{c}^{A} \ ,
\label{COCYCLES_CS_3} \\
\O^{1}_{2}\=\frac{1}{2}f^{ABC}\AA^{A}\AA^{B}\hat{c}^{C}
-\FF^{A}_{mn}e^{m}\h^{n}\AA^{A}-\FF^{A}\hat{c}^{A} \ ,
\label{CHERNTWO} \\
\O^{0}_{3}\=-\frac{1}{6}f^{ABC}\AA^{A}A^{B}\AA^{C}
+\FF^{A}\AA^{A} \ .
\label{CHERNTHREE}
\eea
In particular, expression \equ{CHERNTHREE} gives the familiar
3-dimensional Chern-Simons term.
Finally, let us remark that the
cocycle $\O^{1}_{2}$ of eq.\equ{CHERNTWO},
when referred to dimension $N=2$, reduces to the expression
\eq
\O^{1}_{2}=-(d\AA^{A})c^{A}~~~,~~~{\rm for}~N=2 \ ,
\eqn{GAUGE_ANOMALY_2}
which directly gives the 2-dimensional gauge anomaly.

Let us proceed to give the construction of the
5-dimensional Ashtekar Chern-Simons term.
The descent equations are given by
\bea
&&s\O^{0}_{5}+d\O^{1}_{4}=0 \ ,\non
&&s\O^{1}_{4}+d\O^{2}_{3}=0 \ ,\non
&&s\O^{2}_{3}+d\O^{3}_{2}=0 \ ,\non
&&s\O^{3}_{2}+d\O^{4}_{1}=0 \ ,\non
&&s\O^{4}_{1}+d\O^{5}_{0}=0 \ ,\non
&&s\O^{5}_{0}=0 \ ,
\label{tode5}
\eea
and the cocycles are obtained by using Sorella's method \cite{sor}
\bea
&&\O^{4}_{1}=\d\O^{5}_{0} \ ,\non
&&\O^{3}_{2}=\frac{\d^{2}}{2!}\O^{5}_{0} \ ,\non
&&\O^{2}_{3}=\frac{\d^{3}}{3!}\O^{5}_{0} \ ,\non
&&\O^{1}_{4}=\frac{\d^{4}}{4!}\O^{5}_{0} \ ,\non
&&\O^{0}_{5}=\frac{\d^{5}}{5!}\O^{5}_{0} \ .
\eea
In order to find a solution for the last equation of the tower
given in eq.(\ref{tode5}) we use the redefined Ashtekar ghost
\be
\hat{c}^a_{~b}=A^a_{~bm}\h^{m}+c^a_{~b} \ ,
\ee
which, from eq.(\ref{delta}), transforms as
\be
\d\hat{c}^{a}_{~b}=-A^{a}_{~b} \ .
\ee
We obtain for the 0-form cocycle $\O^{5}_{0}$ in five dimensions
\bea
\O^{5}_{0}\=-\frac{1}{10}\hat{c}^{a}_{~b}
\hat{c}^{b}_{~c}
\hat{c}^{c}_{~d}
\hat{c}^{d}_{~e}
\hat{c}^{e}_{~a}
+\frac{1}{4}F^{a}_{~bmn}\h^{m}\h^{n}
\hat{c}^{b}_{~c}
\hat{c}^{c}_{~d}
\hat{c}^{d}_{~a} \non
\-\frac{1}{4}F^{a}_{~bmn}\h^{m}\h^{n}F^{b}_{~ckl}\h^{k}\h^{l}
\hat{c}^{c}_{~a} \ ,
\eea
which leads to the 5-dimensional Chern-Simons term in
Ashtekar variables
\be
\O^{0}_{5}=\frac{1}{5!}\delta^5\O^5_0=
\frac{1}{10}A^{a}_{~b}A^{b}_{~c}A^{c}_{~d}A^{d}_{~e}A^{e}_{~a}
-\frac{1}{2}F^{a}_{~b}A^{b}_{~c}A^{c}_{~d}A^{d}_{~a}
+F^{a}_{~b}F^{b}_{~c}A^{c}_{~a} \ .
\ee

\section{Appendices:}

Appendix A is devoted to demonstrate the computation of some
commutators involving the tangent space derivative $\6_{a}$.
In appendix B one finds
some relations concerning the determinant of the tetrad
and the $\varepsilon$-tensor.


\section*{A~~~Commutator relations}

\setcounter{equation}{0}
\renewcommand{\theequation}{A.\arabic{equation}}

In order to find the commutator of two tangent space derivatives
$\6_{a}$, we make use of the fact that the usual spacetime
derivatives $\6_{\mu}$ have a vanishing commutator:
\be
[\6_{\mu},\6_{\nu}]=0 \ .
\ee
{}From
\be
\6_{\mu}=e^{m}_{\mu}\6_{m}
\ee
one gets
\bea
[\6_{\mu},\6_{\nu}]=0 \= [e^{m}_{\mu}\6_{m},e^{n}_{\nu}\6_{n}]\non
\= e^{m}_{\mu}e^{n}_{\nu}[\6_{m},\6_{n}]
+e^{m}_{\mu}(\6_{m}e^{n}_{\nu})\6_{n}
-e^{n}_{\nu}(\6_{n}e^{m}_{\mu})\6_{m}\non
\=e^{m}_{\mu}e^{n}_{\nu}[\6_{m},\6_{n}]
+(\6_{\mu}e^{k}_{\nu}-\6_{\nu}e^{k}_{\mu})\6_{k}\non
\=e^{m}_{\mu}e^{n}_{\nu}[\6_{m},\6_{n}]
+(T^{k}_{\mu\nu}-A^{k}_{~n\mu}e^{n}_{\nu}
+A^{k}_{~m\nu}e^{m}_{\mu}
)\6_{k}\non
\=e^{m}_{\mu}e^{n}_{\nu}(T^{k}_{mn}+A^{k}_{~mn}-A^{k}_{~nm}
)\6_{k} \non
\+e^{m}_{\mu}e^{n}_{\nu}[\6_{m},\6_{n}] \ ,
\eea
so that
\be
[\6_{m},\6_{n}]=-(T^{k}_{mn}+A^{k}_{~mn}-A^{k}_{~nm}
)\6_{k} \ .
\ee
\newline
For the commutator of $d$ and $\6_{m}$ we get
\bea
[d,\6_{m}]\=[e^{n}\6_{n},\6_{m}]\non
\=-(\6_{m}e^{k})\6_{k}-e^{n}[\6_{m},\6_{n}]\non
\=-(\6_{m}e^{k})\6_{k}+e^{n}(T^{k}_{mn}+A^{k}_{~mn}
-A^{k}_{~nm})\6_{k} \ ,
\eea
and one has therefore
\be
[d,\6_{m}]=(T^{k}_{mn}e^{n}+A^{k}_{~mn}e^{n}
-A^{k}_{~nm}e^{n} -(\6_{m}e^{k}))\6_{k} \ .
\ee
Analogously, from
\be
[s,\6_{\mu}]=0
\ee
one easily finds
\bea
[s,\6_{m}]\=(\6_{m}\h^{k}-c^{k}_{~m})\6_{k}
+\h^{n}[\6_{m},\6_{n}]\non
\=(\6_{m}\h^{k}-c^{k}_{~m}-T^{k}_{mn}\h^{n}
-A^{k}_{~mn}\h^{n}+A^{k}_{~nm}\h^{n})\6_{k} \ .
\eea


\section*{B~~~Determinant of the vielbein and
              the $\varepsilon$-tensor}

\setcounter{equation}{0}
\renewcommand{\theequation}{B.\arabic{equation}}

The definition of the determinant of the vielbein $e^{a}_{\mu}$
is given by
\bea
e\=det(e^{a}_{\mu})=\frac{1}{4!}\varepsilon_{a_{1}a_{2}a_{3}a_{4}}
\varepsilon^{\mu_{1}\mu_{2}\mu_{3}\mu_{4}}e^{a_{1}}_{\mu_{1}}
e^{a_{2}}_{\mu_{2}}
e^{a_{3}}_{\mu_{3}}e^{a_{N}}_{\mu_{N}} \ .
\eea
One can easily verify that the BRST transformation of $e$ reads
\be
se=-\6_{\l}(\x^{\l}e) \ .
\ee
For the case of $SO(1,3)$ one has
\bea
e^{0}e^{1}e^{2}e^{3}\=\frac{1}{4!}\e_{a_{1}a_{2}a_{3}a_{4}}
e^{a_{1}}e^{a_{2}}e^{a_{3}}e^{a_{4}}\non
\=\frac{1}{4!}\e_{a_{1}a_{2}a_{3}a_{4}}e^{a_{1}}_{\mu_{1}}
e^{a_{2}}_{\mu_{2}}e^{a_{3}}_{\mu_{3}}e^{a_{4}}_{\mu_{4}}
dx^{\mu_{1}}dx^{\mu_{2}}dx^{\mu_{3}}dx^{\mu_{4}}\non
\=\frac{1}{4!}\e_{a_{1}a_{2}a_{3}a_{4}}
\e^{\mu_{1}\mu_{2}\mu_{3}\mu_{4}}
e^{a_{1}}_{\mu_{1}}e^{a_{2}}_{\mu_{2}}e^{a_{2}}_{\mu_{2}}
e^{a_{4}}_{\mu_{4}}
dx^{0}dx^{1}dx^{2}dx^{3}\non
\=ed^{4}\!x=\sqrt{-g}d^{4}\!x \ ,
\eea
where $g$ denotes the determinant of the metric tensor $g_{\mu\nu}$
\be
g=det(g_{\mu\nu}) \ .
\ee
The $\varepsilon$-tensor has the usual norm
\be
\varepsilon_{a_{1}a_{2}a_{3}a_{4}}
\varepsilon^{a_{1}a_{2}a_{3}a_{4}}=-4! \ ,
\ee
and obeys the following relation under partial contraction
of two indices
\be
\varepsilon_{abcd}\varepsilon^{mncd}
=-2(\d^{m}_{a}\d^{n}_{b}-\d^{n}_{a}\d^{m}_{b}) \ ,
\ee
and in general the contraction of two $\varepsilon$-tensors is given
by the determinant of $\d$-tensors in the following way:
\be
\varepsilon_{a_{1}a_{2}a_{3}a_{4}}
\varepsilon^{b_{1}b_{2}b_{3}b_{4}}=-
\left|
\begin{array}{cccc}
\d^{b_{1}}_{a_{1}} & \d^{b_{2}}_{a_{1}}
&\d^{b_{3}}_{a_{1}}& \d^{b_{4}}_{a_{1}}  \\
\d^{b_{1}}_{a_{2}} & \d^{b_{2}}_{a_{2}}
&\d^{b_{3}}_{a_{2}}& \d^{b_{4}}_{a_{2}}  \\
\d^{b_{1}}_{a_{3}} & \d^{b_{2}}_{a_{3}}
&\d^{b_{3}}_{a_{3}}& \d^{b_{4}}_{a_{2}}  \\
\d^{b_{1}}_{a_{4}} & \d^{b_{2}}_{a_{4}}
&\d^{b_{3}}_{a_{4}}& \d^{b_{4}}_{a_{4}}
\end{array}
\right| \ .
\ee


\noindent
{\bf ACKNOWLEDGEMENTS}

We are grateful to all the members of the Institut
f\"{u}r Theoretische Physik
of the Technische Universit\"{a}t Wien for
useful discussions and comments.
One of us (LT) would like to thank Prof. W. Kummer for the
extended hospitality at the institute.


\begin{thebibliography}{99}

\bibitem{ym}  C.N. Yang and R.L. Mills, {\em Phys. Rev.}
              {\bf 96} (1954) 191;\\
              S. Glashow and M. Gell-Mann, {\em Ann. Phys. (N.Y.)}
              {\bf 15} (161) 437;

\bibitem{u}  R. Utiyama, {\em Phys. Rev.} {\bf 101} (1956) 1597;

\bibitem{w}  E. Witten, {\em Comm. Math. Phys.}
             {\bf 117} (1988) 353;\\
             E. Witten, {\em Comm. Math. Phys.}
             {\bf 118} (1988) 411;

\bibitem{bbrt}  D. Birmingham, M. Blau, M. Rakowski and G. Thomson,
                {\em Phys. Rep.} {\bf 209} (1990) 129;

\bibitem{ash}  A. Ashtekar, {\em Phys. Rev. Lett.}
               {\bf 57} (1986) 2247;
               {\em Phys. Rev.} {\bf D36} (1987) 1587;

\bibitem{sp}  J. Samuel and Pram\~ana {\em J. Phys.}
              {\bf 28} (1987) L429;
              {\em Class. Quantum Grav.} {\bf 5} (1988) L123;

\bibitem{rov}  C. Rovelli, {\em Ashtekar formulation of general
               relativity and loop-space non-perturbative quantum
               gravity}; University of TRENTO
               preprint (1991);

\bibitem{cs}  L.N. Chang and C.P. Soo,
              Report No. VPI-THEP 92-5 (unpublished);

\bibitem{cs1}  L.N. Chang and C.P. Soo, {\em Phys. Rev.}
               {\bf 46D} (1992) 4257;

\bibitem{brs}  C. Becchi, A. Rouet and R. Stora,
               {\em Ann. Phys. (N.Y.)} {\bf 98} (1976) 287; \\
               I.W. Tyupin,
               {\em Gauge Invariance in Field Theory and
                Statistical Physics},
                Lebedev Institute preprint FIAN no. {\bf 39} (1975);

\bibitem{wz}  J. Wess and B. Zumino, {\em Phys. Lett.}
              {\bf B37} (1971) 95; \\
              B. Zumino, {\em Chiral anomalies and differential
              geometry}, Les Houches'83,
              eds. B.S. DeWitt and R. Stora, North Holland,
              Amsterdam, 1987;

\bibitem{sor}  S.P. Sorella, {\em Comm. Math. Phys.}
               {\bf 157} (1993) 231;

\bibitem{gms}  E. Guadagnini, N. Maggiore and S.P. Sorella,
               {\em Phys. Lett.} {\bf B255} (1991) 65;\\
               C. Lucchesi, O. Piguet and S.P. Sorella,
               {\em Nucl. Phys.} {\bf B395} (1993) 325;

\bibitem{br}  D. Birmingham and M. Rakowski, {\em Phys. Lett.}
              {\bf B269} (1991) 103;
              {\em Phys. Lett.} {\bf B272} (1991) 217;
              {\em Phys. Lett.} {\bf B275} (1992) 289;
              {\em Phys. Lett.} {\bf B289} (1992) 271;

\bibitem{wss}  M. Werneck de Oliveira, M. Schweda and S.P. Sorella,
               {\em Phys. Lett.} {\bf B315} (1993) 93;

\bibitem{boss}  A. Boresch, M. Schweda and S.P. Sorella,
                {\em Vector supersymmetry of the superstring in
                super-Beltrami parametrization} CERN-TH-7110/93,
                preprint TU Wien Ref.TUW 93-27;

\bibitem{mss}  O. Moritsch, M. Schweda and S.P. Sorella,
               {\em Class. Quantum Grav.} {\bf 11} (1994) 1225;

\bibitem{diss} O. Moritsch and M. Schweda, {\em Helv. Phys. Acta}
               {\bf 67} (1994) 289;

\bibitem{dix}  J. Dixon, {\em Comm. Math. Phys.}
               {\bf 139} (1991) 495;

\bibitem{ban}  G. Bandelloni, {\em Nuovo Cimento}
               {\bf 88A} (1985) 1;
               {\em Nuovo Cimento} {\bf 88A} (1985) 31;

\bibitem{st}  S.P. Sorella and L. T\u ataru, {\em Phys. Lett.}
              {\bf B324} (1994) 351;


\bibitem{tm1}  J. Thierry-Mieg, {\em J. Math. Phys.}
              {\bf 21} (1980) 2834;
              {\em Nuovo Cimento} {\bf 56A} (1980) 396;
              {\em Phys. Lett.} {\bf B147} (1984) 430;

\bibitem{sto}  R. Stora, {\em Algebraic structure and topological
               origin of anomalies}, Carg\`ese '83, eds.
               G. t'Hooft et. al., Plenum Press, New York, 1987;

\bibitem{witten}  L. Alvarez-Gaum{\'e} and E. Witten,
                  \np{{\bf B234}}{83}{269}

\bibitem{ba1}  L. Baulieu, \np{{\bf B241}}{84}{557} \\
               L. Baulieu and J. Thierry-Mieg,
               \pl{{\bf B145}}{84}{53}

\bibitem{dvio}  M. Dubois-Violette, M. Talon and C.M. Viallet,
                     {\it Comm. Math. Phys.} {\bf 102} (1985) 105;
                     \pl{{\bf B158}}{85}{231}
                     \aihp{{\bf 44}}{86}{103} \\
  M. Dubois-Violette, M. Henneaux, M. Talon and C.M. Viallet,
                       {\it Phys. Lett.} {\bf B267} (1991) 81;\\
  M. Henneaux, \cmp{{\bf 140}}{91}{1}\\
  M. Dubois-Violette, M. Henneaux, M. Talon and C.M. Viallet,
              {\it Phys. Lett.} {\bf B289} (1992) 361;

\bibitem{gins}  L. Alvarez-Gaum{\'e} and P. Ginsparg,
                    \annp{{\bf 161}}{85}{423}

\bibitem{ton1} L. Bonora, P. Pasti and M. Tonin,
                 \jmp{{\bf 27}}{86}{2259}

\bibitem{stora} F. Langouche, T. Sch\"ucker and R. Stora,
                \pl{{\bf B145}}{84}{342} \\
                R. Stora , {\em Algebraic structure of
                chiral anomalies};
                lecture given at the GIFT seminar,
                1-8 june 1985, Jaca, Spain;
                preprint LAPP-TH-143; \\
                J. Manes, R. Stora and B. Zumino,
                \cmp{{\bf 102}}{85}{157} \\
                T. Sch\"ucker, \cmp{{\bf 109}}{87}{167}

\bibitem{brandt}  F. Brandt, N. Dragon and M. Kreuzer,
                  \np{{\bf B340}}{90}{187}

\bibitem{tm2}  J. Thierry-Mieg, {\em Classical geometrical
               interpretation of ghost fields and anomalies
               in Yang-Mills theory and quantum gravity},
               Symposium on Anomalies, Geometry, Topology,
               pag. 239, edited by W.A. Bardeen
               and A.R. White, World Scientific, 1985;

\bibitem{bb}  L. Baulieu and M. Bellon, {\em Phys. Lett.}
              {\bf B161} 96;
              {\em Nucl. Phys.} {\bf B266} (1986) 75;

\bibitem{bmsstz}  P. Blaga, O. Moritsch, M. Schweda, T. Sommer,
                  L. T\u ataru and H. Zerrouki,
                  {\em Algebraic structure of gravity in
                  Ashtekar variables}, preprint
                  TU Wien Ref.TUW 94-15;

\bibitem{cjd}  R. Capovila, T. Jacobson and J. Dell,
               {\em Phys. Rev. Lett.} {\bf 63} (1989) 2325;

\bibitem{mss1}  O. Moritsch, M. Schweda and T. Sommer,
               {\em Yang-Mills gauge anomalies in the presence of
               gravity with torsion},
               preprint TU Wien Ref.TUW 94-13;

\bibitem{ws}  M. Werneck de Oliveira and S.P. Sorella,
              {\em Int. Journ. Mod. Phys.} {\bf A9} (1994) 2979;

\bibitem{fb}  F. Brandt, {\em Structure of BRS-Invariant Local
              Functionals}, Ref.NIKHEF-H 93-21;

\end{thebibliography}
\end{document}